\documentclass[prb,reprint,superscriptaddress,amsmath,amssymb]{revtex4-1}
\usepackage{graphicx}
\usepackage{dcolumn}
\usepackage{bm}
\usepackage{color}

\begin{document}

\title{$^{23}$Na NMR study of sodium order in Na$_{x}$CoO$_{2}$ with 22~K N\'eel temperature}

\author{H.~Alloul}
\affiliation{Laboratoire de Physique des Solides, CNRS UMR 8502, Universit\'e Paris-Sud, 91405 Orsay, France, EU}
\author{I.R.~Mukhamedshin}
\email{Irek.Mukhamedshin@ksu.ru}
\affiliation{Institute of Physics, Kazan Federal University, 420008 Kazan, Russia} \affiliation{Laboratoire de Physique des Solides, CNRS UMR 8502, Universit\'e Paris-Sud, 91405 Orsay, France, EU}
\author{A.V.~Dooglav}
\affiliation{Institute of Physics, Kazan Federal University, 420008 Kazan, Russia} \affiliation{Laboratoire de Physique des Solides, CNRS UMR 8502, Universit\'e Paris-Sud, 91405 Orsay, France, EU}
\author{Ya.V.~Dmitriev}
\affiliation{Institute of Physics, Kazan Federal University, 420008 Kazan, Russia} \affiliation{Laboratoire de Physique des Solides, CNRS UMR 8502, Universit\'e Paris-Sud, 91405 Orsay, France, EU}
\author{V.-C.~Ciomaga}
\affiliation{Institut de Chimie Mol\'eculaire et des Mat\'eriaux d'Orsay, CNRS UMR 8182, Universit\'e Paris-Sud, 91405 Orsay, France, EU}
\author{L.~Pinsard-Gaudart}
\affiliation{Institut de Chimie Mol\'eculaire et des Mat\'eriaux d'Orsay, CNRS UMR 8182, Universit\'e Paris-Sud, 91405 Orsay, France, EU}
\author{G.~Collin}
\affiliation{Laboratoire de Physique des Solides, CNRS UMR 8502, Universit\'e Paris-Sud, 91405 Orsay, France, EU}

\begin{abstract}
We report a systematic study of the $c$ lattice parameter in the Na$_{x}$CoO$%
_{2}$ phases versus Na content $x>0.5$, in which sodium always displays
ordered arrangements. This allows us to single out the first phase which
exhibits an antiferromagnetic order at a N\'{e}el temperature $T_{N}=$22~K which is found to occur for $x\approx 0.77(1)$. Pure samples of this phase have been studied both as aligned powders and single crystals. They exhibit identical $^{23}$Na NMR spectra in which three sets of Na sites could be
fully resolved, and are found to display $T$ dependencies of their NMR
shifts which scale with each other. This allows us to establish that the $T$
variation of the shifts is due to the paramagnetism of the Co sites with
formal charge state larger than 3$^{+}$. The existence of a sodium site with
axial charge symmetry and the intensity ratio between the sets of $^{23}$Na
lines permits us to reveal that the two-dimensional structure of the Na order corresponds to 10 Na sites on top of a 13 Co sites unit cell, that is with $%
x=10/13\approx 0.77$. This structure fits with that determined from local
density calculations and involves triangles of 3 Na sites located on top of
Co sites (so called Na1 sites). The associated ordering of the Na vacancies
is quite distinct from that found for $x<0.75$.
\end{abstract}



\maketitle

\section{Introduction}

Since the discovery of high thermo-electric power\cite{TerasakiTEP} and
superconductivity\cite{TakadaNature} in layered cobaltates Na$_{x}$CoO$_{2}$%
, extended efforts have been made in order to understand the importance of
electronic correlations in their metallic and magnetic properties. While the
$x=1$ compound is a band insulator,\cite{LangNa1,MHJulienNa1} a slight
reduction of Na content down to $x\approx 0.75$ results in various
antiferromagnetic (AF) phases with distinct N\'{e}el temperatures detected
by local probe measurements such as $\mu $SR \cite{SugiyamaPRL2004,Mendels05}%
, NQR/NMR \cite{EPL2008,MHJulien075,JETP4phasesNQR} or by thermodynamic
properties \cite{Foo,SchulzeBatlogg}. Neutron scattering data established
that the AF state is of $A$-type that is ferromagnetic in the CoO$_{2}$
planes and AF between planes as long as $x\gtrsim 0.75$.\cite%
{BayrakciNa082,BoothroydNa075} NMR studies established that similar
Curie-Weiss paramagnetism occurs above 100~K down to $x=2/3$, though ordered
magnetism does not occur below $x$ of the order 0.75.\cite{EPL2008} The
study of the spin dynamics by spin lattice measurements in NMR also has
evidenced that ferromagnetic correlations are still dominant in plane \cite%
{EPL2008}, but switch abruptly to AF correlations in plane for $x\leq 0.62$.%
\cite{LangNFD}

While many experiments and theoretical calculations have considered that the
Co magnetism is uniform, it has been evidenced by NMR that Na displays an
atomic ordering associated with Co charge disproportionation in the planes.%
\cite{CoPaper,ImaiPRL1,Ray,Gavilano1,Ishida07,EPL2008,MHJulien075,H67_CoNMR} Such Na ordered atomic structures have been observed by TEM,\cite{Zandbergen} neutrons,\cite{Roger} and x-rays.\cite{TaiwanPRB2009,H67NQRprb,FouryPRB} In Na$_{1}$CoO$_{2}$ the single Na site is located on top of the center of a Co triangle, in a configuration usually called Na2. Numerical simulations \cite{Roger} and electronic structure calculations \cite{Hinuma} suggest that the Na2 vacancies formed for $x<1$, are ordered in the Na plane. These vacancies have a tendency toward clustering and, depending on $x$, these clusters induce altogether the appearance of isolated Na1 sites (on top of a Co) in divacancy clusters or of trimers of Na1 sites in trivacancy clusters.

Of course the incidence of the structural order on the charge
disproportionation and on the physical properties is still an important
pending question. However, the experimental situation that prevails so far
is quite unusual in solid state physics, as most experiments do not permit
altogether to establish reliably the relation between the local order
proposed, the actual Na content and the local magnetic properties of the
studied samples. We have demonstrated that NMR/NQR is a powerful technique
allowing us to establish this correlation, and applied it to the specific $%
x=2/3 $ phase.\cite{EPL2009} There we found ordered divacancies resulting in isolated Na1 sites in the Na plane. They are accompanied by the differentiation of Co$^{3+}$ non magnetic sites, on top and below the Na1, with respect to Co sites with a formal valence $\approx 3.44$ located on a kagom\'{e} substructure of the triangular cobalt lattice. The local magnetic properties of these cobalt sites could be studied in great detail.\cite{H67_CoNMR} Theoretical work suggests that the electronic correlations are enhanced for this $x=2/3$ composition,\cite{LFrank_PRL107} and that this Na content is indeed the turning point between AF and ferromagnetic in-plane correlations.\cite{BoehnkeLechermann}

To expand our work toward higher Na contents for which the Na order
might have a prominent incidence on the magnetic 3D order, we have selected
to study here the magnetic phase with a well defined $T_{N}=22$~K which we
had detected quite early by $\mu $SR.\cite{Mendels05} This appears to be one
of the most stable magnetic phases, which has been found by many authors.
However a large controversy still exists on the actual Na content of
this phase which some have found near the composition $x\approx
0.75$\cite{BoothroydNa075,Mendels05,MHJulien075,PrbTN22KSpecificHeat} while
others reported a composition of $x=0.82$.\cite{TaiwanPRB2009} This latter
estimate of Na content introduced some confusion in the scientific community
and required considering complicated staging structures with different Na
contents on alternating planes, a situation which has been therefore
proposed for most Na concentrations.\cite{TaiwanPRB2009}

Here we present experimental work that resolves these data differences but
also allows us to characterize the in plane Na atomic order in this $%
T_{N}=22 $~K phase. We could synthesize both a single phase powder sample of
this phase that has been aligned in the NMR field, and single
crystals in which this phase could be isolated. The x-ray data analysis
and the $^{23}$Na NMR results allow us to confirm on secure physical grounds
our calibration scale, which has been recently confirmed by electrochemistry.%
\cite{NatureMat2010} We demonstrate that the $T_{N}=$22~K phase corresponds
to a Na content of $x \approx 0.77$ and matches the simple planar Na order proposed in Ref.~\onlinecite{Hinuma} with 13 Co per unit cell containing only Na trivacancies (or Na1 trimers). The $^{23}$Na NMR data are not
sufficient so far to permit us to fully resolve the 3D stacking of this unit
cell. But they allow us to conclude that this stacking does not require any
complex staging, and that the concentration $x$, which can be deduced from
our $c(x)$ curve permits one to reconcile most existing data.

\section{Sample synthesis and characterization}

We discuss here the preparation of the sample powders, and give evidence
that we could produce single phase samples. The analysis of the x-ray data
allows us then to determine a calibration curve for the variation of the $c$
axis parameter versus $x$ and report the existing composition gaps.
Preparation of samples with a floating zone technique which usually allows
to synthesize homogeneous single crystals will then be discussed and it will
be shown that in such conditions multiphase samples with various N\'{e}el
temperatures are usually produced.

\subsection{Powder samples}

We have detailed in former publications the method used to synthesize powder
samples suitable for NMR experiments in the range of concentrations between
0.45 and $\sim $0.78.\cite{LangNFD,EPL2008} There we have shown that x-ray
diffraction allowed us to select single phase samples, that is phases with a
well defined Na ordering. Indeed, the systematic study of the powder x-ray
spectra for various nominal Na contents allowed us to separate pure phases
from multiphase samples with distinct $c$ axis parameters. We took
systematic x-ray spectra for samples with $x>1/2$ for increasing Na content,
by steps of 0.3\% in Na content. We could then track all the hexagonal two
layers $P2$ phases ($P6_{3}/mmc$, n$^{\circ }$166) which can be quenched at room $T$ up to the $T_{N}=$22~K phase.

This preparation procedure also allowed us to produce the distorted $O3$
rhombohedral (monoclinic) phase conventionally named three-layers which cannot be prepared in oxygen atmosphere at high temperature (850-900$^{\circ }$C).
They are only obtained either by Na de-intercalation starting from $x=1$ or
by direct synthesis in argon atmosphere at 600-700$^{\circ }$C depending on
composition.

\begin{figure}[tbp]
\includegraphics[width=1.0\linewidth]{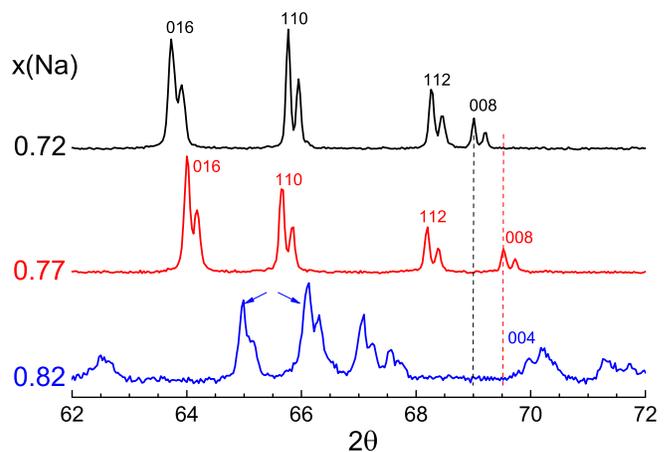}
\caption{(Color online) Part of the x-ray powder spectra between 62 and 72$%
^{\circ }$ reflection angles for $x=0.72$, the $T_{N}=22~$K phase and the
higher $x$ monoclinic phase with $T_{N}=29$~K. For any $0.72<x<0.77$, the
spectra are superpositions of the two spectra of the pure phases. One can
see that in these pure phases the simple observation of the (008) Bragg
peaks permits ensuring the absence of phase mixing. The 0.82 phase is monoclinic $C2/m$ (n$^{\circ }$12, one-layer) structure with a splitted 110 reflection (2 arrows in the figure) and a 004 equivalent to the hexagonal (two-layers) 008 reflection. This structure results from a strong shear distortion of the rhombohedral $R-3m$ (three-layers) modification of the Na-Co-O2 packing. (The double peak structure corresponds here to the two Bragg peaks associated with the Cu K$\protect\alpha _{1}$ and K$\protect\alpha _{2}$ radiations).}
\label{FigXray22K}
\end{figure}

We illustrate that point by displaying the powder diffraction Bragg
peaks shown in Fig.~\ref{FigXray22K} for the pure $T_{N}=22$~K phase and the
closest phase with lower Na content. For intermediate concentrations between
these, we always find mixed-phase samples with the two end phases. The
diffraction peaks of the monoclinic phase which occurs in the same range of
angles as that of the (008) for the hexagonal phase is also displayed in
Fig.~\ref{FigXray22K} for the pure $T_{N}=29$~K phase detected initially by $%
\mu $SR in our research group.\cite{Mendels05} The diffraction angles for
our pure phases are quite similar to those reported by Shu \textit{et al} on
single crystal multiphase samples (see their Fig.~2 in Ref.~%
\onlinecite{TaiwanPRB2009}). So we certainly identified the same ordered
phases, but we have synthesized them as pure phases. We could
not synthesize in powder form the pure magnetic phase with $T_{N}=9$~K which
is located between the $T_{N}=22$~K and $T_{N}=29$~K phases, although we got
mixed phase samples. We shall see later that we also found this $T_{N}=9$~K phase embedded in single crystals.

The actual Na content of these phases can then be monitored by the $c$ axis
parameter value, which could be accurately obtained in our case by full
Rietveld analysis of the diffractions on a range of angles $2\theta
=10^{\circ }-130^{\circ }$ (for Cu K$_{\alpha }$). But estimates of the actual Na content require a determination of the calibration curve $c(x)$ on such
single phase samples. We determined accurately three points:

(i) the $x=1/2$ phase for which the structure is well established and $%
c=11.1310(3)$~\AA ;

(ii) the $x=2/3$ phase for which we have determined the structure by NMR/NQR,
\cite{EPL2009} which has then been fully confirmed by x-ray Rietveld
analysis with $c=$10.9383(3)~\AA ;\cite{H67NQRprb}

(iii) the $x=0.71$ pure phase, which has a characteristic NQR spectrum,\cite%
{JETP4phasesNQR} quite distinct from that for $x=2/3$, and could be
synthesized with full thermogravimetric control from the constituents Na$%
_{2} $CO$_{3}$ and Co$_{3}$O$_{4}$ without mass loss. For this phase $c=$10.8874(3)~\AA .

\begin{figure}[tbp]
\includegraphics[width=1.0\linewidth]{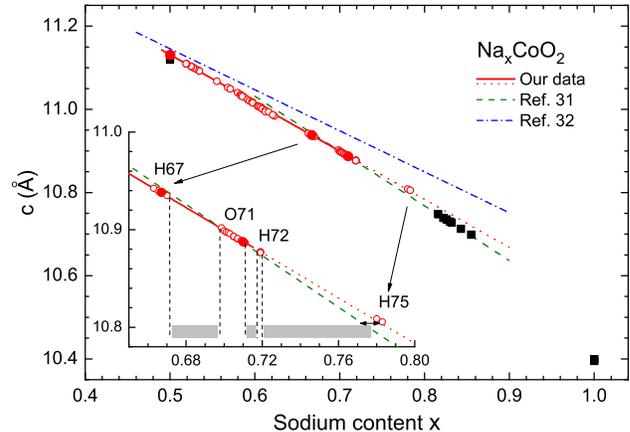}
\caption{(Color online) Calibration curve for the $c$ axis parameter versus
Na content for the $P2$ hexagonal two layers phases (full line) based on the
three data points plotted as filled circles (see text). We reported on this full line the $c$ parameter values (empty circles) obtained for all pure phase
samples we could synthesize. Intervals with no data point marked by grey
bars correspond to composition gaps. Linear extrapolation toward the data
points corresponding to samples for which $T_{N}=22$~K suggest $x\approx 0.78
$ for that phase. The sample compositions studied in Ref.~
\onlinecite{EPL2008} are reported as H67, O71, H72, and H75. The filled
squares are data for the monoclinic $O3$ three-layers phases (see text). They
apparently lie on a slightly distinct $c(x)$ line starting from $x=1/2$. The
$c(x)$ curve reported in Ref.~\onlinecite{NatureMat2010} (dashed, green) agrees well with our data, while that reported in Ref.~\onlinecite{TaiwanPRL2008}
(dash-dotted, blue) departs markedly.}
\label{FigCaxisVsXb}
\end{figure}

We notice, as shown in Fig.~\ref{FigCaxisVsXb}, that these three points
define a linear relation between $x=0.5$ and $x=0.71$. This plot allowed us
to determine the concentration $x$ of the pure phases which could be
stabilized at room $T$ in the absence of air contamination for all
intermediate concentrations. We have reported in Fig.~\ref{FigCaxisVsXb}
their $c$ parameters and delineate there the composition gaps for which only
mixed phase samples can be produced. This allows us to locate the sample
compositions we have been studying so far by NQR/NMR up to $x=0.72$.\cite%
{LangNFD,EPL2008} It is clear that the large composition gap which occurs
above $x=0.72$, and the structural change which occurs above allowed us to
identify reliably and reproducibly the $T_{N}=22$~K phase. Indeed, the single
phase sample with the (008) diffraction peak located at $\theta \approx
69.5^{\circ }$ at room temperature was found to display a single magnetic
transition detected at $T_{N}=22$~K by $\mu $SR experiments.\cite{Mendels05}
Notice that, for the powder samples of this phase we did not succeed in
indexing the weak superstructure or incommensurate satellites due to Na
order, contrary to the case of all the samples with $x\leq 0.72$.

The recent study \cite{NatureMat2010} of the phase diagram done by
electrochemical control of the Na content allowed us to verify our $x=2/3$
structure, and permitted us to confirm a phase diagram in overall quite good
agreement with ours. There, the difference between the 0.71 and 0.72 phases
quite well evidenced from NMR/NQR data \cite{EPL2008, JETP4phasesNQR} could
not however be resolved. Their $c(x)$ calibration curve, in good agreement
with ours, is displayed on Fig.~\ref{FigCaxisVsXb}, together with the quite
distinct one determined in Refs.~\cite{TaiwanPRB2009,TaiwanPRL2008} by the
inductively coupled plasma method and electron probe analysis. \footnote{%
This has lead them to associate their four magnetic phases with $x=$ 0.763,
0.820, 0.833 and 0.859. We do not think that any of the existing data could
determine the concentration of Na in a pure phase sample to better than 0.5\%%
}

For the $T_{N}=22$~K composition, assuming that a linear extrapolation of
our $c(x)$ curve applies, one would deduce $x\approx 0.78$, slightly above the
value $0.75$ that has been assumed by many including us and in total
disagreement with the $x=0.820$ value of Shu \textit{et al.} \cite%
{TaiwanPRB2009} The results of Ref.~\onlinecite{NatureMat2010} indicate that
the $c(x)$ curve might, however, not be linear up to the highest Na contents
and might bend slightly down so that the $T_{N}=22$~K phase might correspond
to $0.76<x<0.78$.

Let us point out that powder samples with composition larger than 0.78 could only be synthesized in the three layer monoclinic structure O3. The corresponding $c$ axis values projected on the hexagonal cell $c$ axis also have been reported in Fig.~\ref{FigCaxisVsXb}. The $x$ values are however reliable as the samples were synthesized at relatively low $T$ with therefore negligible Na loss.

The linear variation that appears to apply from the monoclinic $x=1/2$
phase up to about $x=0.82$ is only slightly different from that for the
hexagonal phases and does not extrapolate at the $c$ axis value for the $x=1$
rhombohedral composition, which is a completely different structural
modification with a large composition gap above $x=0.86$.

For all the high Na content phases with $x>0.72$ the samples evolved if submitted to any humid air as had been monitored from detailed
evolution of the NQR spectra, which allowed to evidence the Na loss.\cite%
{JETP4phasesNQR} NQR and zero field NMR experiments at 4.2~K on a non
oriented fraction of a $T_{N}=22$~K phase sample allowed us to ensure that
the sample was not contaminated at the 10\% level by any paramagnetic phase
with lower $x$, for which the NQR signals had been perfectly identified.\cite%
{JETP4phasesNQR}

After alignment in an applied field, with the procedure described in Ref.~%
\onlinecite{Egorov}, the very sample on which most of the NMR experiments
had been done had been kept at liquid N$_{2}$ temperature and only heated
to 200~K during experimental runs and never taken back to room $T$. NMR
experiments allowed us to confirm that the phase content of this sample did
not evolve significantly during the alignment process, as detailed in
Appendix~\ref{AppendixNMRControl}.

\subsection{Single crystal samples}

Synthesis of single crystal samples has been performed in an image furnace
by a floating zone technique. Pieces which appeared as single crystals could
be cleaved in some parts of the 11~cm long synthesized bars. To permit rf
field penetration for the NMR experiments on samples exceeding 100~mg, we
sliced these samples into thin pieces of about 200~$\mu $m thickness and
piled those upon each other with teflon tape separations.

\begin{figure}[tbp]
\includegraphics[width=1.0\linewidth]{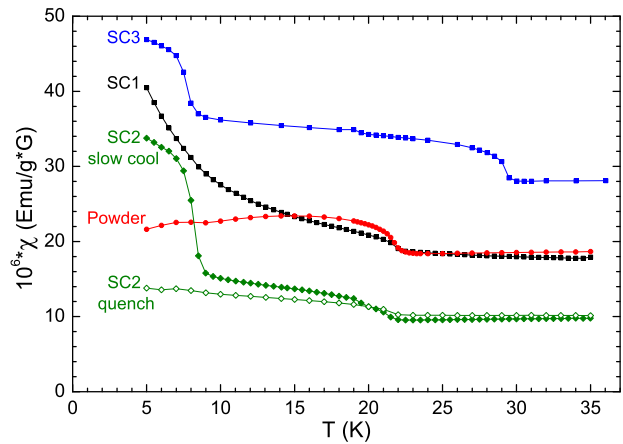}
\caption{(Color online) Magnetization data taken with a SQUID magnetometer
on a non aligned powder sample and on three Na$_{x}$CoO$_{2}$ single
crystal samples with high sodium content. The data were taken after slow
cooling (5~K/min) from room $T$ to 5~K in a field of 100~Oe. The single
phase powder sample with x$\approx 0.78$ is found to exhibit a single
magnetic transition at 22~K. The crystals data were taken for $H\perp c$.
SC3 exhibits magnetic transitions at 29~K and 9~K, while SC2 exhibits
transitions at 22~K and 9~K. If quenched down to LN$_{2}$ temperature,
the 9~K transitions are found to be suppressed as evidenced here for
instance for the SC2 sample.}
\label{FigNa077Squid}
\end{figure}

Three such samples named SC1, SC2 and SC3 have been used and their magnetic
properties were controlled first by SQUID measurements which established
that the Na content was not uniform in those macroscopic samples. One could
easily see in Fig.~\ref{FigNa077Squid} the signatures of three magnetic
transitions at 9~K, 22~K and 29~K, which differed in the three samples.
While SC1 and SC2 exhibited initially transitions at 9~K and 22~K, SC3
exhibits the 9~K and 29~K transitions. So these transitions are
characteristic of independent Na ordered phases and the samples are mixed
phases presumably of adjacent phases in the phase diagram.

So, such an image furnace synthesis does not allow a full control of the Na
content, which was found dependent on the excess Na used in the starting
materials and of the Na loss which depends critically on the growth
conditions. Although the samples appear \textit{as single crystals for the ordering of the CoO$_{2}$ lamellas}, the homogeneity of Na content is not realized in such macroscopic samples, which often display multiphase content. This was similarly found by many others as well.\cite%
{SchulzeBatlogg,TaiwanPRB2009} As will be shown hereafter, $^{23}$Na NMR data
helped us to better characterize the crystallinity of these samples and the
phase content.

The 22~K and 29~K phases were known and isolated already in powder samples
for the $\mu $SR experiments,\cite{Mendels05} while, as said above, we never
were able to synthesize powder samples displaying a majority phase with $%
T_{N}=$9~K. Also for the 22~K and 29~K phases SQUID magnetic transition
signals were not found to depend on the cooling process of the sample. On
the contrary, as can be seen in Fig.~\ref{FigNa077Squid}, the SQUID
transition signal of the 9~K phase could be depressed by fast cooling below
200~K, which indicates that some specific Na ordering has to be established
by slow cooling to stabilize the $T_{N}=9$~K phase.\cite{SchulzeBatlogg}

We expected that such bulk samples would be more stable than the powders but
this happened to be only partly the case, as we found that they did still
evolve slowly in time when kept in air. The sample SC1 which initially was a
mixture of 22~K and 9~K phase did lose progressively the 9~K phase signal
and finally had only the 22~K phase left. As shown in Appendix~\ref%
{AppendixSampleEvolution}, we could even control partly this type of
evolution by heat treatment.

Our observations essentially agree with those done independently by specific
heat and magnetization measurements,\cite{SchulzeBatlogg} and with the
hierarchy of N\'{e}el temperatures with increasing Na content reported in Ref.~\onlinecite{TaiwanPRB2009} on samples which are apparently similar
mixed phase single crystals. Then, contrary to our ceramic synthesis
done in the solid state for the powder samples, it does not seem easy to
synthesize directly phase pure single crystal samples with controlled Na
content. This might explain the difficulty to fix the Na content by chemical
analysis which measure the total Na/Co ratio, and not that actually involved
in a specific phase. In view of this evolution of the single crystals at
room $T$, we decided then to keep our three samples in liquid nitrogen to
avoid further evolutions of their Na contents. This was of particular
importance for the SC1 sample which had progressively expelled its 9~K phase
and had become a nearly pure 22~K phase sample, as will be confirmed below
by $^{23}$Na NMR. As will be detailed hereafter, these data will furthermore
allow us to get independent confirmation of the validity of our $c(x)$
calibration curve.

Let us point out that although in Ref.~\onlinecite{TaiwanPRL2008} Na content
calibrations do not match ours in Fig.~\ref{FigCaxisVsXb}, the present
observations indicate unambiguously that we have been performing
measurements on samples with the same characteristic structural and physical
properties as those studied by Shu \textit{et al.}\cite{TaiwanPRB2009}

\section{$^{23}$N\lowercase{a} NMR data}

In the first NMR work done in our group on cobaltates we immediately
showed that $^{23}$Na NMR is a powerful tool, which allowed us to establish the existence of Na atomic order.\cite{NaPaper} As $^{23}$Na has a spin $I=3/2$,
the NMR spectrum for a single Na site displays a central transition and two
satellites disposed symmetrically with respect to the central line. While the shift of the central line signal is governed by the magnetism of the
near-neighbor Co sites, the distance $\Delta \nu $ between the satellites
is linked to the quadrupole frequency $\nu _{Q}$ associated with the
magnitude of the electric field gradient (EFG) at the Na site, which is
governed by the distribution of ionic and electronic charges around the Na.

\subsection{NMR spectra for $H\parallel c$}

\begin{figure*}[tbp]
\includegraphics[width=0.8\linewidth]{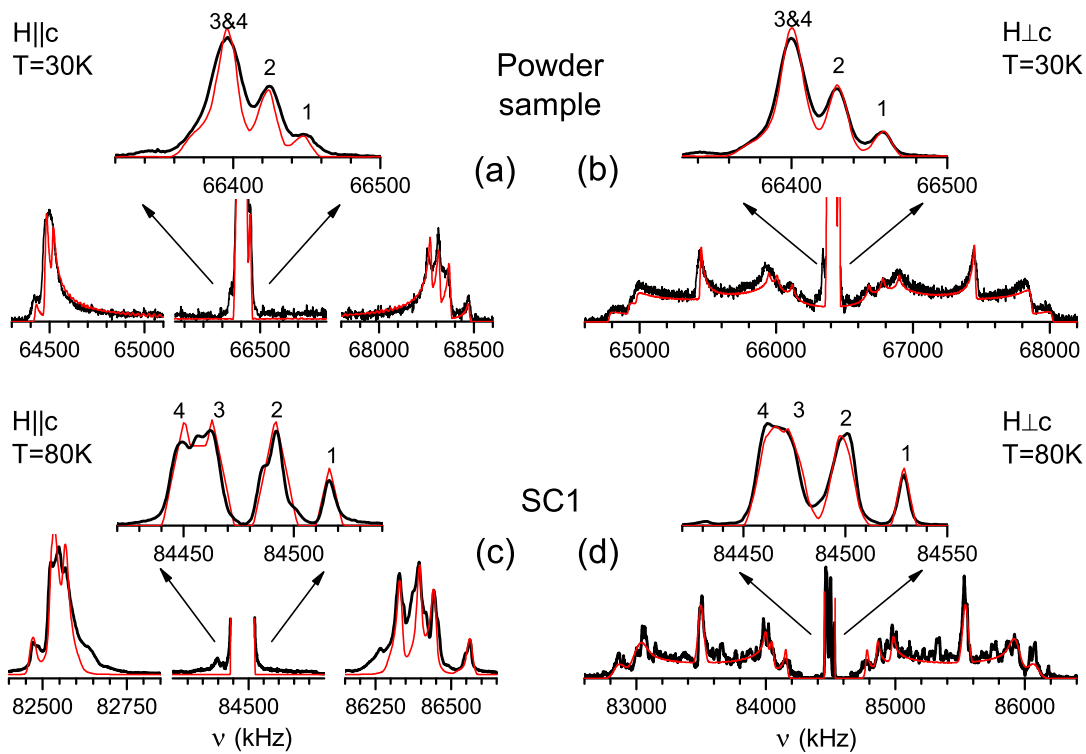}
\caption{(Color online) $^{23}$Na NMR spectra taken in the paramagnetic
state on the Na$_{x}$CoO$_{2}$ single phase samples with $T_{N}=$22~K. The
spectra for $H\parallel c$ and $H\perp c$, displayed on the top (a) and (b)
panels for the powder sample were taken in $H\simeq $5.9~T. In the (c) and
(d) panels data for the SC1 sample taken for $H\simeq $ 7.5~T are displayed. Notice that in all panels the frequency sweeps of the full spectrum have been cut between the right and left quadrupole satellites and the central line. An expanded view of the latter is shown on the top of each panel and the Na NMR sites are labeled there from 1 to 4 (reported as Na(1) to Na(4) in the text). The computer simulations of the NMR spectra with the parameters given in Table~\ref{tab:NaSimParam} are shown as thin red lines (see text).}
\label{FigNa077Spectra}
\end{figure*}

So, as in the $x=2/3$ phase,\cite{NaPaper} we find here that, with an
external field applied along the $c$ axis of the powder or crystal samples
the $^{23}$Na NMR spectrum reflects above $T_{N}$ the diverse Na sites
pertaining to the studied phase. The NMR spectra of the oriented powder
sample of the $T_{N}=$22~K phase displayed three well resolved lines on the
central transition and on the high frequency satellites, but which overlap
for the low frequency satellites as can be seen in Fig.~\ref{FigNa077Spectra}%
(a). The three sets of central lines detected are similar to those found
in the Ref.~\onlinecite{MHJulien075} on a single crystal of this $T_{N}=$22~K phase. In both cases the less shifted line is the broadest so that one can anticipate that it corresponds to a superposition of unresolved lines with slightly differing shifts and $\nu_{Q}$ values.

The $^{23}$Na NMR spectrum of the ``purified'' SC1 sample in which the 22~K
phase is dominant displays identical features with a slightly better
resolution and slightly narrower satellite lines - see Fig.~\ref%
{FigNa077Spectra}(c). This indicates that the $c$ axis orientation is better
defined there, as ensured by the easy cleavage, while perfect alignment of
the $c$ axis is harder to achieve in powder samples. The differentiation of
the high frequency quadrupole satellites in the SC1 sample is better and
ensures the occurrence of at least four Na sites distinguished by their
distinct $\nu _{Q}$ values.

Let us point out that the $^{23}$Na NMR spectra are quite distinct in the 9~K phase as will be shown in Appendix~\ref{AppendixNMR9K22K}. It is quite important now to consider data for $H\perp c$ to be able to sort out the respective local properties of the set of resolved Na lines.

\subsection{NMR spectra for $H\perp c$: local symmetry of the Na sites}

Taking the data for $H\perp c$ is essential to get good indications on the
local symmetry of the sites. Let us recall that the distance $\Delta \nu $
between the two satellite lines which correspond for the $^{23}$Na to the $\frac{3}{2}\leftrightarrow \frac{1}{2}$ and $-\frac{1}{2}\leftrightarrow -\frac{3}{2}$ transitions depends on the orientation of the external field with respect to the $Z$ principal axis of the EFG tensor which lies in the $c$ axis direction in these lamellar systems.\cite{NaPaper,EPL2008} $\Delta \nu $ can be expressed as:%
\cite{Abragam}
\begin{equation}
\Delta \nu =\nu _{Q}(3\cos ^{2}\theta -1+\eta \sin ^{2}\theta
\cos 2\varphi ),
\end{equation}%
where $\eta =(V_{XX}-V_{YY})/V_{ZZ}$ is the asymmetry parameter of the EFG
tensor and $\theta $ and $\varphi $ the spherical angular coordinates of the
field. In the case of powder crystallite samples, the observed spectrum for $%
H\perp c$ is the superposition of the spectra obtained for all possible
orientations of the $a-b$ plane. A remarkable case, very easy to identify, is
that for sites with axial symmetry for which $\eta =0$, so that $\Delta \nu $
is unique whatever the orientation of the field in the $a-b$ plane and
exactly occurs at $\nu _{Q}$, that is half the value $\Delta \nu =$ $2\nu
_{Q}\ $ obtained for $H\parallel c$. For non zero values of $\eta $ the
quadrupolar satellites display double horn spectra in the powder spectrum as
seen in Fig.~\ref{FigNa077Spectra}(b), the extension of the double horn
being then determined by the $\eta $ value.

The inspection of the spectrum of Fig.~\ref{FigNa077Spectra}(b) quite easily
allows us to identify the existence of an axial site, which is found to
correspond to that which displays the largest shift of the central line in
both $H\parallel c$ and $H\perp c$ spectra. From the width of the line in
the $a-sb$ direction and the measured values of $\Delta \nu $ in the two field
directions, an upper limit of $0.02$ can be put on its asymmetry parameter.
As will be seen later \textit{the identification of this axial site is a
quite essential element} which will allow us hereafter to resolve the Na in
plane atomic order. Notice that, as can be seen in the $c$ axis spectrum,
the intensity of this line is much weaker than that of the other Na sites.
We shall come back on these intensity determinations later.

The other feature that is easily seen when comparing the spectra for $%
H\perp c$, Fig.~\ref{FigNa077Spectra}(b) for the powder sample and Fig.~\ref%
{FigNa077Spectra}(d) for SC1, is the strong analogy between the spectra. In
both samples the $\eta =0$ axial site is detected as well as identical
double horns. This allows us to confirm that the SC1 is not a single crystal
in the $a-b$ plane and contains at least as many twin boundaries and more
probably a mosaic of crystallites of in plane Na order, so that the SC1
spectrum appears quite analogous to the powder sample spectrum. However the
number of crystallites is not large enough to span the random angular  distribution of the powder sample, this being seen as an ``additional noise'' in the $H \perp c$ spectrum of SC1 as compared to the powder sample case. For both samples the simulations of the spectra allow us to establish that the three non axial Na sites display large $\eta $ values between 0.5 and 0.7 as listed in Table \ref{tab:NaSimParam}.

\subsection{Intensities of the various Na lines and site occupancies}

It is quite easy to see in the expanded central line spectra, for both $%
H\parallel c$ and $H\perp c$, and for both samples that the most shifted $%
\eta =0$ line is the less intense, while the middle line has lower intensity
than the less shifted one, that results from the overlap of at least two
signals. To get accurate relative intensities one needs to take into account
the decay time of the spin-echo signal which differs from site to site, and
to control very accurately the reproducibility of the results. Also, as we
cannot guarantee the phase purity to better than 90\% in these cobaltate
samples, weak signal components of impurity phases might superimpose on
those of the dominant phase and influence the data. It is then quite
difficult to obtain a determination of the relative intensities on a single
spectrum with an accuracy better than 10\%.

The $\eta =0$ line being the most shifted hardly overlaps with the other
ones and its intensity can be estimated relative to the total intensity. We
find that the statistical number for this ratio, taken on a large series
of spectra can be secured to (9.5$\pm 0.5)\%$ on both the powder and SC1
samples. This would indicate that this Na(1) NMR spectrum corresponds to one Na site over 10 or 11. If part of the less shifted side of the spectra includes that of some weak intensity impurity phases, the actual intensity ratio for the $\eta =0$ line could be rather 10\% so that it would correspond rather to 1
site on a total of 10. The other intensity ratio which can be secured is Na(2)/Na(1)$=3\pm 0.2$, which fully excludes considering 4 sodium sites for the Na(2) NMR spectrum.

As for the less shifted part of the spectrum which involves at least two
sets of lines Na(3) and Na(4), which are poorly resolved, we have refrained
from determining the relative intensities of these unresolved lines and have done only careful determinations of the intensity ratio between Na(3\&4) and
Na(2). Here again the statistical determination of this ratio could be
evaluated to be $1.95\pm 0.1$. indicating that the less shifted lines
correspond to twice the number of sites of the intermediate one. Overall,
the most realistic number of sites which might be retained would be 6/3/1
with increasing NMR shifts, and could hardly be 7/3/1 or 5/3/1, as one would
expect then an intensity ratio of 2.33 or 1.66 between the two intense sets
of lines, significantly out of the experimental error bar.

\subsection{Full analysis of the Na NMR spectra and $T$ dependence of the
NMR shifts}

Overall it has been quite possible to simulate the two sets of spectra using
the magnetic shift ($K_{xy},K_{z}$) and quadrupole parameters ($\nu _{Q}$, $\eta $) given in Table~\ref{tab:NaSimParam} for the four detected sites, as can be seen in Fig.~\ref{FigNa077Spectra}. Though, the respective intensities
3/3/3/1 used cannot be ascertained solely by these fits, especially for the
two components of the lowest shift lines. But one might notice that the simulations are similarly good for the powder and SC1 samples.

\begin{table}[tbp]
\caption{Parameters used for computer simulations of the $^{23}$Na NMR
spectra shown in Fig.~\protect\ref{FigNa077Spectra}. The principal
components of the magnetic shift tensor $K_{\protect\alpha }$ are given in
the principal axes ($X,Y,Z$) of the EFG tensor and measured relative to the $^{23}$Na NMR in NaCl solution. The actual error bar on  $\nu_Q$ is about 5~kHz and for the magnetic shifts is about 0.005\%.}
\label{tab:NaSimParam}%
\begin{ruledtabular}
\begin{tabular}{cccccc}
& NMR site & Na(1) & Na(2) & Na(3) & Na(4) \\
\hline
& Weight & 1 & 3 & 3 & 3 \\
\hline
& $\nu_Q$ (MHz) & 2.026 & 1.94 & 1.92 & 1.88 \\
Powder & $\eta$ & 0 & 0.5 & 0.7 & 0.58 \\
30~K & $K_{XY}$ (\%) & 0.156 & 0.115 & 0.077 & 0.07 \\
& $K_{Z}$ (\%) & 0.156 & 0.118 & 0.075 & 0.073 \\
\hline
& $\nu_Q$ (MHz) & 2.044 & 1.951 & 1.932 & 1.88 \\
SC1 & $\eta$ & 0 & 0.5 & 0.68 & 0.57 \\
80~K & $K_{XY}$ (\%) & 0.1685 & 0.136 & 0.080 & 0.0923 \\
& $K_{Z}$ (\%) & 0.164 & 0.136 & 0.099 & 0.0853 \\
\end{tabular}
\end{ruledtabular}
\end{table}

\begin{figure}[tbp]
\includegraphics[width=1.0\linewidth]{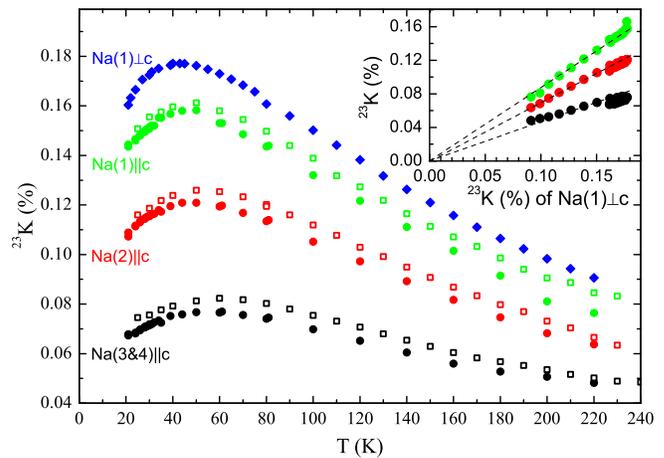}
\caption{(Color online) $T$ variation of the NMR shifts for the $^{23}$Na
lines shown in Fig.~\protect\ref{FigNa077Spectra} for $H\parallel c$. The
full circles are the SC1 sample data. Here, the average values for Na(3) and
Na(4) data have been taken to permit comparison with the powder sample
(empty circles) for which those are not resolved. Data for Na(1) in SC1 for $%
H\perp c$ are shown for comparison as well by full diamonds. The SC1 data
for $H\parallel c$ are reported in the insert versus that for $H\perp c$ for
Na(1), showing that a single $T$ dependence dominates the shift for all
sites in both directions.}
\label{FigNa077Shifts}
\end{figure}

We did follow the $T$ dependence of the shifts $K_{Z}(T)$ of the various
lines in the spectra, which are plotted in Fig.~\ref{FigNa077Shifts} and are
quite similar for the powder and SC1 samples. We also plotted for reference
the shift of the outer Na(1) line for $H\perp c$ that is $K_{XY}(T)$. In all
cases the shifts display a maximum at about 40K as was already reported in
Ref.~\onlinecite{EPL2008} for the mean shift of the spectrum. We found that $%
K_{XY}(T)$ and $K_{Z}(T)$ are nearly identical for the Na(1) lines but also
for the other ones, which confirms the isotropy of the spin susceptibility
found from SQUID data. Furthermore, as was found as well for the $x=2/3$
phase,\cite{NaPaper} the shifts for the various lines scale with each other
as shown in the inset of Fig.~\ref{FigNa077Shifts}. This indicates that, as
concluded as well for the $x=2/3$ phase, the various Na sites in the
structure are detecting the magnetism of their neighbouring Co sites through
slightly distinct hyperfine paths, and that a single $T$ dependence
dominates the physical properties.

\section{2D Structure of the N\lowercase{a} planes}

\begin{figure}[tbp]
\includegraphics[width=1.0\linewidth]{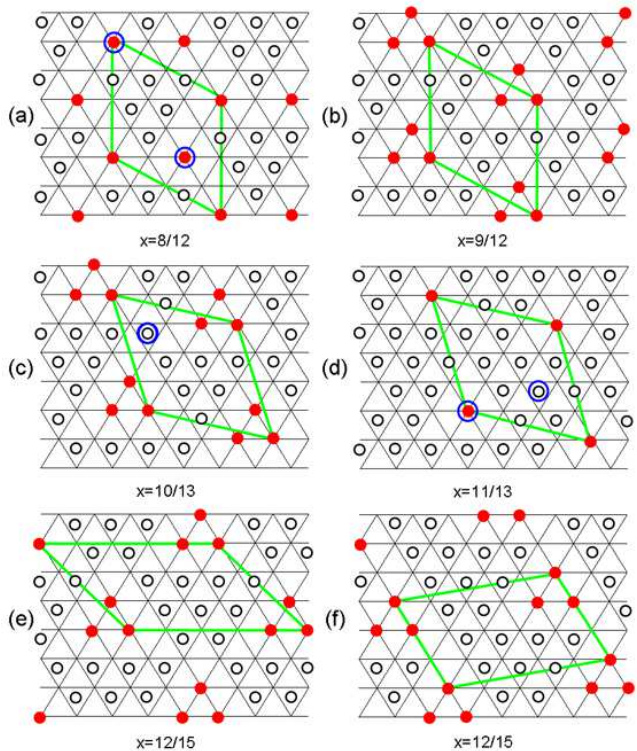}
\caption{(Color online) 2D Na unit cell (a) for $x=2/3$\protect\cite%
{EPL2009}, calculated in Ref.~\onlinecite{Hinuma} from GGA$+U$ for $%
x=9/12$ (b) and from GGA calculations for $x=10/13$ (c). The Na1 sites
(filled red dots) and Na2 sites (empty black circles) are differentiated, and
represented above the triangular lattice of Co sites (not reported) which
are located at all intersections of lines. The $x=11/13$ structure (d)
has been proposed in Ref.~\onlinecite{TaiwanPRB2009} to complement (c) in a
compatible 3D stacking for $x=$0.82. The two structures (e) and (f) have
been proposed for $x=12/15$ in Ref.~\onlinecite{Roger,Morris}. Apart (a) and
(d), which consist of divacancies around isolated Na1, all the other
structures correspond to an ordering of Na2 trivacancies around a trimer of
Na1 sites. In all cases a 2D unit cell is drawn. The only Na sites with
three fold symmetry which should give an axial EFG in the $^{23}$Na NMR in
these structures are distinguished by an extra blue circle.}
\label{FigAFStructures}
\end{figure}

The present results give important elements to decide about the Na order and
content in this $T_{N}$=22~K phase. We have therefore considered the Na
orderings proposed by various authors for $0.75\leq x\leq 0.80$. The minimal
energy structures computed for this $x$ range by Hinuma \textit{et al.} by
GGA calculations or GGA$+U$ calculations correspond to Na concentrations $%
9/12=0.75$ (Fig.~\ref{FigAFStructures}(b)) and $10/13 \approx 0.77$ (Fig.~%
\ref{FigAFStructures}(c)).\cite{Hinuma} These structures contain
trivacancies, which correspond to a triangle of six vacant Na2 sites filled
by a trimer of Na1 sites at the center. The stability of such a cluster of
vacancies has also been emphasized by Roger \textit{et al.} from simulations
done for higher sodium concentrations.\cite{Roger} From their neutron
scattering experiments these authors later identified two distinct
structures which also contain this Na1 trimer, and apparently occur for $%
x=12/15=0.8$ - see Figs.~\ref{FigAFStructures}(e)(f).\cite{Roger,Morris} We
show as well in Fig.~\ref{FigAFStructures}(d) the unit cell with a single
divacancy, proposed by Shu \textit{et al.} for $x=11/13\approx $0.846. They
suggested that two such planar structures are stacked with one $x=10/13
\approx 0.77$ layer (Fig.~\ref{FigAFStructures}(c)) to obtain a structure
corresponding to $x=0.82$, the Na content they proposed for the $T_{N}=22$~K
phase.\cite{TaiwanPRB2009}

The structure of Fig.~\ref{FigAFStructures}(c) corresponds to a 13 cobalt planar unit cell with $x=0.77$ and four Na sites with intensities 1/3/3/3. There the singly occupied Na site is an Na2 site with perfect threefold symmetry
which can be assigned to the $\eta =0$ low intensity Na(1) site in our
spectra. The three other have the same occupancies, so that our spectra
agree with this structure if we consider that the lowest frequency line
combines two lines corresponding to 3 Na sites each. \textit{The simulations done in Fig.~\ref{FigNa077Spectra} with four sites with occupancies 1/3/3/3 are therefore fully compatible with this structure.}

The two $x=12/15$ structures of Figs.~\ref{FigAFStructures}(e)(f)
proposed in Refs.~\onlinecite{Roger,Morris} both exhibit one Na2 site
with axial symmetry for their first neighbours for (e), or for both first
and second neighbours for (f), but not at the third neighbour level, so that
one expects a small $\eta $ value but not $\eta =0$. If we tentatively
assign the experimental $\eta =0$ line to this site, its intensity would be
1/12 of the Na NMR intensity, somewhat smaller than the actually measured
intensity within our experimental accuracy. Furthermore these structures
differentiate six Na sites with intensities 1/2/2/2/2/3 for Fig.~\ref%
{FigAFStructures}(e) and even more sites with intensities 1/1/2/2/2/2/2 for Fig.~\ref{FigAFStructures}(f). As the Na(1) and Na(2) lines correspond
respectively to 1 and 3 sites, the 8 remaining sodium sites would correspond
to the Na(3\&4), so one would expect for the intensity ratio of the two
lower shift NMR lines Na(3\&4)/Na(2) = $8/3 \approx 2.66$ which is largely
in excess of our data of $1.95\pm 0.1$.

As for the $x=11/13 \approx 0.846$ structure proposed in Ref.~%
\onlinecite{TaiwanPRB2009}, shown in Fig.~\ref{FigAFStructures}(d), one can see that it has one Na1 site and one Na2 site which both display a perfect axial symmetry, so that these two distinct sites would correspond to $\eta =0$. If two such planes were stacked with the $x=10/13$ structure of Fig.~\ref{FigAFStructures}(c), as proposed in the Ref.~\onlinecite{TaiwanPRB2009}, we would expect three distinct locally axial sites, which is not compatible with our data for which a single site with $\eta =0$ is resolved. Also that would result in a quite complicated spectrum
with up to 7 other different Na sites, totally incompatible with the
relatively simple spectrum detected hereabove.

So, for the $T_{N}=22$~K phase, the structure which is compatible with all
the features of our $^{23}$Na NMR data is that of Fig.~\ref{FigAFStructures}%
(c) with $x=10/13 \approx 0.77$. However in the $^{23}$Na NMR spectra, we are not able in this unit cell to assign the Na(2)-Na(4) signals, with 3 sodium positions each, which do not have axial symmetry. Such a correspondence could only be established by considering the 3D stacking of the Na planes, taking into account differentiation of the Co sites - see Section \ref{Section3DStruct22Kphase}.

If one now considers the synchrotron x-ray data reported in Ref.~\onlinecite{TaiwanPRB2009} on the samples with $T_{N}=$22~K, with the same $c$ axis parameter as ours, as well as those on the 9~K and 29~K phase, they find that they do all correspond to a lattice with a 13 cobalt unit cell, that is $\sqrt{13}\ast \sqrt{13}$. Their structural result agrees then with the fact that the cells of Fig.~\ref{FigAFStructures}(b)(e)(f) with either 12 or 15 Co per unit cell cannot explain our $^{23}$Na NMR data. On the contrary the 13Co unit cell illustrated in Fig.~\ref{FigAFStructures}(c) proposed in Ref.~\onlinecite{Hinuma} from GGA calculations, for $x=10/13 \approx 0.77$, \textit{is fully compatible with all the data obtained on the} $T_{N}$=22~K \textit{phase}, that is:

- the local symmetries and signal intensities of the Na sites detected in
our $^{23}$Na NMR data;

-the size $\sqrt{13}\ast \sqrt{13}$ of the unit cell obtained by single
crystal x-ray data;

- the Na content estimated from our $c(x)$ calibration curve and that of
Ref.~\onlinecite{NatureMat2010}.

\section{3D structure of the 22~K phase ?}
\label{Section3DStruct22Kphase}

So far we have only done an analysis based on a single layer of Na between
CoO$_{2}$ layers. In real crystals one has to stack the Na layers to define
a 3D structure. We further know that in all phases with $x>0.5$ the charges
of the Co sites are differentiated. For $2/3<x<0.8$ we have demonstrated the
existence of 20 to 40\% of Co$^{3+}$ in our samples.\cite{EPL2008} This has
been found as well by others.\cite{MHJulien075,ImaiPRL1} So this
differentiation of the Co charges depends of the stacking of the Na layers
above and below the CoO$_{2}$ layer.

We have studied that in detail in the case of $x=2/3$ for which the Na layer
contains two identical Na1 sites and two distinct Na2 sites with multiplicity
3 as recalled in Fig.~\ref{FigAFStructures}(a). Due to the high symmetry of
that structure, the stacking of the layers does not differentiate the Na
sites of the individual layers and the full crystal structure only retains the 3 Na sites of the single layer.

\begin{figure}[tbp]
\includegraphics[width=1.0\linewidth]{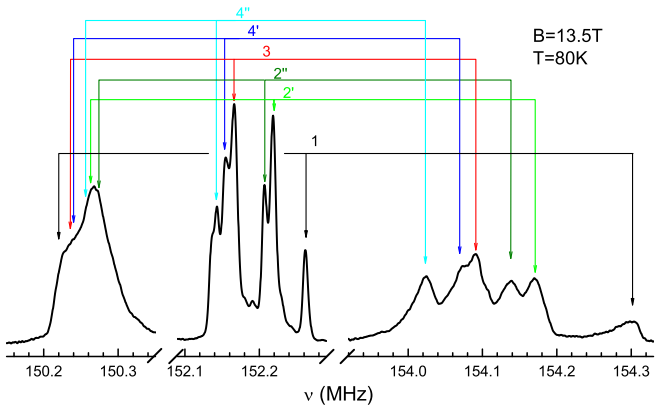}
\caption{(Color online) $^{23}$Na NMR spectrum of the SC1 sample, for $H=13.5
$~T. In this field a better resolution than in Fig.~\protect\ref%
{FigNa077Spectra}(c) is achieved for $H \parallel c$. Here, the Na(2) and
Na(4) lines exhibit weak splittings which are seen both on the central
transition and the high frequency satellite. The correspondence between
central lines and satellite is shown by arrows.}
\label{FigCr1HFNaNMR}
\end{figure}

But with the low symmetry of the unit cell for x=10/13, the 3D stacking of
this layer could differentiate the crystallographic sites of the four Na
sites distinguished in the 2D structure. Weak splittings can indeed be guessed in the 7.5~Tesla spectra of Fig.~\ref{FigNa077Spectra}. We have then
taken data in a higher field of 13.5~Tesla to try to better resolve such
splittings on the SC1 sample in which the spectral resolution is better than
on the powder sample. On the spectrum displayed in Fig.~\ref{FigCr1HFNaNMR},
we can indeed see that the central line and right satellites are better
resolved in this applied field, and we distinguish weak subsplittings of the
Na(2) and Na(4) lines into lines labeled as Na(2$^\prime$), Na(2$^{\prime \prime}$), Na(4$^\prime$) and Na(4$^{\prime \prime}$). Similar subsplitting could be seen as well on the central line data in Ref.~\onlinecite{MHJulien075}, so that seems to be sample independent and characteristic feature for this phase with $T_{N}=22$~K.

If one may consider that the 3D stacking establishes a perfect crystalline
order, the rough analysis of the NMR intensities of the central line and
high frequency satellites would indicate that both Na(2) and Na(4) split
into two lines with multiplicities of (1:2). Therefore the 3D structure
could correspond to 6 Na sites with respective Na occupancies of
1:2:3/1:2/1, with increasing Na NMR shift.

We attempted to use these differences between Na sites to guess the actual
stacking of the 10/13 structure. We have to recall that this stacking
should reveal the ordering of the Co charges differentiated in each plane
depending of the Na neighbours of each Co site. This would then in turn
imply a differentiation of the Na sites depending of their Co neighbours in
the 3D structure. This was successfully done in the $x=2/3$ phase, where
the symmetry of the 2D unit cell stacking was so high that it did not result
in a further differentiation of the three planar Na sites. The situation
might be more complicated in the present case as we have to determine how
the Na1 trimers are positioned on the two Na layers below and above the CoO$%
_{2}$ layers. The number of possibilities is much larger in this case than for the $x=2/3$ phase. Obviously, even in the latter case, the knowledge of the charge differentiation of the Co was necessary to solve the structure. We are
therefore presently undertaking a detailed study of the $^{59}$Co NMR
spectra, which might help us to finalize such an analysis and to give us
indications on the Co charge disproportion in this phase.

\section{Conclusion}

Here we have done a systematic investigation of the stable Na ordered
phases, and the composition gaps occuring in Na$_{x}$CoO$_{2}$ cobaltates
in the range $0.65<x<0.90$. We focused our attention on the magnetic
phases, the first one occuring for $x=0.77(1)$ with $T_{N}=22$~K, and did
evidence that both our powder and single crystal samples give phases with
magnetic properties identical to those reported in the literature. This
allowed us then to confirm that the $c$ axis versus $x$ calibration curve we
have established for long is in good agreement with that proposed by
producing similar phases by electrochemical reduction of Na content.\cite{NatureMat2010} Our results totally dismiss the $c(x)$ curve reported by Shu \textit{et al.} solely by chemical analysis of the Na content in their single crystal samples.\cite{TaiwanPRL2008} Those methods always give a larger estimate of the actual Na content of the dominant phase, inasmuch as the extra Na required to grow the single crystals remains in the sample as chemical impurities.\cite{Bordeaux2009}

More importantly, we have performed a specific $^{23}$Na NMR study of pure
phase powder and single crystal samples of the $T_{N}=22$~K phase and
could identify 4 distinct Na sites, one of which displaying an axial
symmetry. This allowed us to identify the 13 Co unit cell of this
phase, which contains 10 Na (hence $x=10/13 \approx 0.77$), and which agrees with the size of the cell proposed by Shu \textit{et al.} from their synchrotron x-ray single crystal data.\cite{TaiwanPRB2009} The agreement of this value of $x$  with that deduced from our $c(x)$ calibration does not call for any complex staging for this composition. The 3D unit cell should then result from a stacking of 2D $x=10/13$ Na unit cells. We could not however anticipate so far this 3D stacking solely from $^{23}$Na NMR data.

The important new feature of these results is that the unit cell established
here experimentally contains a cluster of three Na2 vacancies surrounding a triangle of Na1 sites which has been suggested for long to occur by various simulations for larger Na contents.\cite{Roger,Hinuma} On the contrary for all the ordered phases for $x<0.75$ that we have studied so far Na2 divacancies
order around isolated Na1 sites. In the $x=2/3$ phase the Na1 sites were
found directly on top of  the Co$^{3+}$ sites with fully filled  $t_{2g}$
orbitals, and this appears to be the case for all $x<0.75$. In the
present case the disproportionation of Co states might be quite distinct.
So one might anticipate that the A type AF phases which only occur above $x=0.75$ can be driven by this distinct order of the Na sites. In such a
case the two significant boundaries anticipated in the Ref.~\onlinecite{Hinuma,MengHinuma} in the phase diagram of Na cobaltates appear to be confirmed experimentally. Divacancies order in rows below $x=0.65$,\cite{LangNFD}, display 2D order for $x=2/3$ which transform into zigzag chains up to $x=0.72$.\cite{EPL2009} They disappear then for $x=10/13$ and are replaced by trivacancies which presumably persist until $x=0.86$ as suggested by GGA calculations done in Refs.~\onlinecite{Hinuma,MengHinuma}. Further detailed Co NMR data should help us to determine the 3D order of the $x=10/13$ phase and to identify how the charge disproportionation is connected with the Na order in these AF phases.

An important question remains as it is not clear so far whether the Co
charge disproportionation which is not considered in the GGA calculations \cite{Hinuma,MengHinuma} is driven by the Na order or whether it is an intrinsic property of the CoO$_{2}$ planes. The kagome type of charge structure could be anticipated from electronic structure calculations,\cite{KoshibaeMaekawa} is even reinforced by electronic correlations\cite{LFrank_PRL107} and happens to match
perfectly the Na order. It is however harder to imagine here that the 13
Co unit cell would be a naturally stable configuration of the charge
disproportionation of the CoO$_{2}$ plane!

\section{Acknowledgments}

We would like to thank here F.~Bert, J.~Bobroff and P.~Mendels
for their help on the experimental NMR techniques and for constant
interest and stimulating discussions. We also thank here Z.Z.~Li for his help for the x-ray characterization of the single crystal samples. We acknowledge the Marie Curie program for the fellowship given for the internship of Y.~Dmitriev, and ``Triangle de la Physique'' for partial support provided for an image furnace at Institut de Chimie Mol\'eculaire et des Mat\'eriaux d'Orsay (ICMMO), and for supporting the visits to Orsay of I.R.M. and A.V.D. Also I.R.M. and A.V.D. thanks for partial support of this work the Russian Foundation for Basic Research (Project No.10-02-01005a), and the Ministry of Education and Science of the Russian Federation (Project No.2010-218-01-192 and Budget Theme No.12-24).

\appendix

\section{NMR control of the powder sample after alignment}

\label{AppendixNMRControl}

\begin{figure}[tbp]
\includegraphics[width=1.0\linewidth]{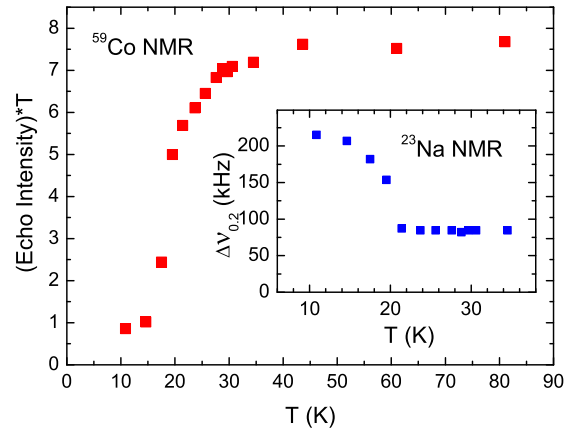}
\caption{(Color online) Variation with temperature of $^{59}$Co NMR
intensity at the position of central line NMR in the paramagnetic phase, for
the $T_{N}=$22~K phase powder sample. The growth of the internal field in
the AF phase shifts the $^{59}$Co NMR out of the observation range, hence
the low signal intensity below $T_{N}$. In the inset, the sharpness of the
magnetic transition at 22~K is also shown by the increase of the $^{23}$Na
NMR linewidth $\Delta H_{0.2}$ (see text)).}
\label{FigNa077Inten}
\end{figure}

To test whether the phase content of the powder sample evolved during the
alignment process, and that the $T_{N}=22$~K remained the majority phase, we
monitored in Fig.~\ref{FigNa077Inten} the evolution of the $^{59}$Co NMR
signal for $T<22$~K. For $H\parallel c$, the NMR signal intensity at the
central line position exhibits an abrupt total loss indicating that the
internal field which develops below $T_{N}$ completely wipes out the $^{59}$%
Co NMR signal from our observation window. Notice that the weak progressive
loss seen above 22~K is due to the slowing down of the magnetic fluctuations
when approaching $T_{N}$, which shortens the transverse relaxation time $%
T_{2}$ and induces the weak reduction of the detected spin echo signal.

On the contrary, in the AF A type phase, in which the Co layers are
ferromagnetically ordered in plane, the two cobalt planes adjacent to the Na
layer induce opposite local fields on the $^{23}$Na nuclei. Those therefore
should only sense weak extra local fields in the magnetic phase, determined
by the actual 3D stacking of the Na planes. We indeed found that the $^{23}$%
Na NMR spectrum only weakly broadens below 22~K. As the central line spectra
of Fig.~\ref{FigNa077Spectra}(a) are somewhat complicated, we just monitored
this broadening by the distance $\Delta H_{0.2}$ between the two extreme
points in the spectrum for which the signal intensity is 20\% of its maximum
intensity, shown in the inset of Fig.~\ref{FigNa077Inten}. The purity of the
22~K phase could be further confirmed by the absence of any anomaly in the $%
^{23}$Na NMR spectra in the paramagnetic phase which could be associated
with the nearby phases with $T_{N}=$9~K and 29~K which were detected in
single crystal (see Appendix~\ref{AppendixNMR9K22K}).

\section{Sample evolutions upon Na extraction}

\label{AppendixSampleEvolution}

\begin{figure}[tp]
\includegraphics[width=1.0\linewidth]{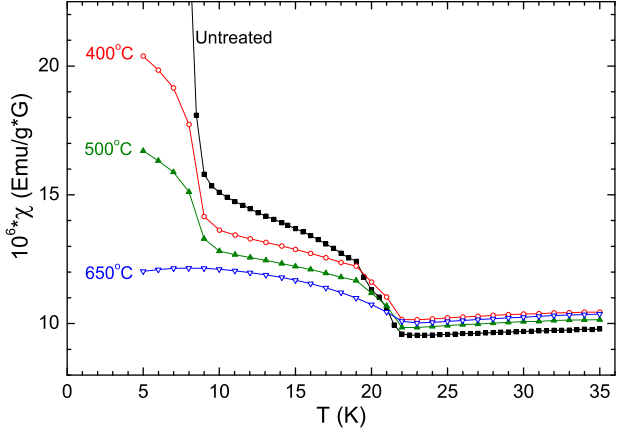}
\caption{(Color online) SQUID magnetization data taken after slow cooling
for a slice of the SC2 sample. The magnitude of the 9~K transition decreases
when the sample is heat treated in argon gas at increasing temperatures,
while the 22~K transition is only slightly modified. Such a heat treatment
therefore expels progressively the Na from the 9~K phase.}
\label{FigNa077HeatTreat}
\end{figure}

Here we studied the evolution with heat treatment of a small slice of the
SC2 sample which contained both the 9~K and 22~K phases. We studied the
progressive variation of the SQUID magnetization data after curing the
sample for a few hours under argon atmosphere at increasing temperatures. As
can be seen in Fig.~\ref{FigNa077HeatTreat} the magnetization signal below
9~K decreased progressively, quite faster than that measured between 9~K and
22~K, which is dominated by the contribution of the 22~K phase. This allowed
us to demonstrate that the 9~K phase corresponds to a slightly higher Na
content than the 22~K phase, as was also evidenced in the Ref.~\onlinecite{TaiwanPRB2009}.

Let us point out that the magnitude of the magnetization below $T_{N}$ is
driven by magnetic anisotropy parameters not well controlled so far.
Therefore, from SQUID data we only get some indication of the actual phases
present in the samples, and their transformations, but that does not allow
us to determine the fraction of phase contents in the sample.

Finally the comparison of our x-ray data, of SQUID data taken on the three
single crystal samples, shown in Fig.~\ref{FigNa077Squid}, their
evolution with heat treatment, allowed us to establish that the magnetic
phases correspond to increasing Na content for the 22~K-9~K-29~K phases, in
quite good agreement with the data of Shu \textit{et al.} on the same phases.\cite{TaiwanPRB2009}

\section{Comparison of the 9~K and 22~K phases}

\label{AppendixNMR9K22K}

\begin{figure}[tbp]
\includegraphics[width=1.0\linewidth]{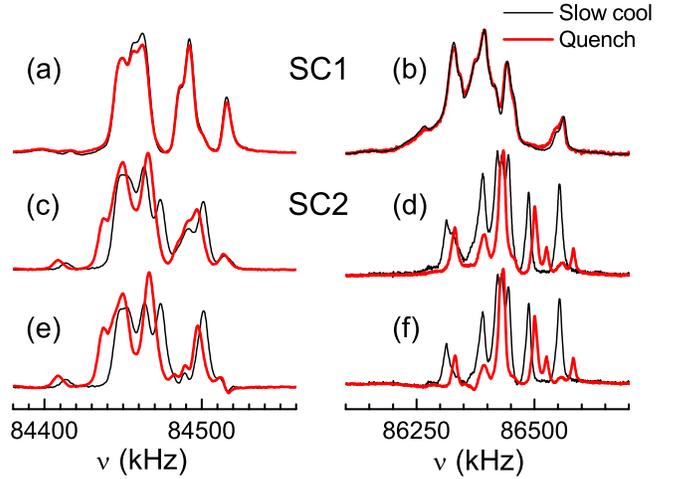}
\caption{(Color online) The $H\parallel c$ $^{23}$Na NMR spectrum, taken at
80~K, of the $T_{N}=22$~K pure phase of the SC1 sample for the central
line (a) and the right satellite (b) does not vary with the rate of cooling
of the sample. For the SC2 sample which is a mixture of the $T_{N}=22$~K and
$T_{N}=9$~K phases, the spectra of both the central line (c) and the right
satellite (d) are quite dependent of the rate of cooling through 200~K.
The variation is associated with the contributions of the phase with $T_{N}=9$~K when slowly cooled. The spectra of the latter, deduced as detailed in the text, are shown in panels (e)(f).}
\label{FigSC1SC2Heattreat}
\end{figure}

We have seen that the $T_{N}=9$~K phase although hardly synthesized as an
isolated phase so far, has been however detected by many authors. It
represents indeed a large part of our SC2 sample. We therefore compared
the $^{23}$Na NMR signal of the SC2 with that of the SC1, which is a pure $%
T_{N}=22$~K phase sample. We did not see any modification of the SQUID data
of the $T_{N}=22$~K SC1 sample with the cooling process and it can be seen
as well on Fig.~\ref{FigSC1SC2Heattreat}(a,b) that the $T$=80~K NMR spectra in
the paramagnetic phase are also totally independent of the cooling process.

This contrasts with our data for the SC2 sample, for which we did find
radical changes of the $^{23}$Na NMR spectrum in the paramagnetic phase when
quenching the sample as displayed in Fig.~\ref{FigSC1SC2Heattreat}(c,d). So the phase which displays a $T_{N}=9$~K transition upon slow cooling displays
distinct Na orders depending of the cooling process. The total independence
of the SC1 signal upon the cooling process is then a further proof of the
absence of $T_{N}=9$~K phase in this sample.

All these results are consistent with the observation by Schulze \textit{et
al.} that the $T_{N}=9$~K phase was obtained by slowly cooling their
samples.\cite{SchulzeBatlogg} Our NMR results further establish that this phase is an independent phase present in multiphase samples, and not a transformation of the $T_{N}=22$~K phase during the cooling process.

Let us notice that Morris \textit{et al.} ~\cite{Roger,Morris} have taken
neutron scattering data on a sample of similar Na content in which they
found a structural change occuring at 240~K. They attribute this to a shear
occuring on the ordered Na structure which shifts from the unit cell of Fig.~
\ref{FigAFStructures}(e) to that of Fig.~\ref{FigAFStructures}(f). In view
of the change observed by slow cooling of the 9~K phase, it might correspond
to the phase studied in Refs.~\onlinecite{Roger,Morris} although one would still need a complete set of data to establish the structure they proposed.

It is then easy to notice in Fig.~\ref{FigSC1SC2Heattreat} that the outer
Na(1) line is characteristic of the 22~K phase spectrum, so from its
intensity in the SC2 spectrum, we could estimate that the fraction of 22~K
phase in the SC2 sample is about 25\%. We could then obtain by substraction
the $^{23}$Na NMR spectra of Fig.~\ref{FigSC1SC2Heattreat}(e)(f), which are
respectively characteristic of the Na order in the slowly cooled $T_{N}=9$~K
phase, and of its modification upon quenching.

\begin{figure}[tbp]
\includegraphics[width=1.0\linewidth]{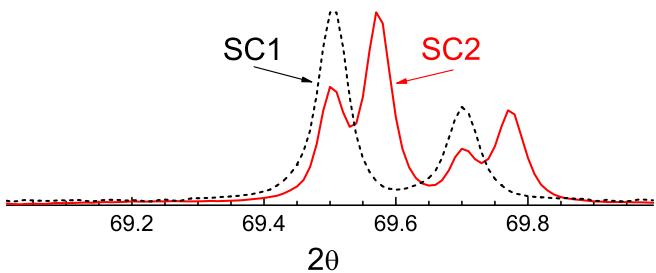}
\caption{(Color online) Part of the x-ray powder spectra with (008) reflections for the $T_{N}=22$~K pure phase of the SC1 sample (black dashed line) and SC2 sample which contains both $T_{N}=22$~K and $T_{N}=9$~K phases (red solid line). (The double peak structure corresponds here to the two Bragg peaks associated with the Cu K$\protect\alpha _{1}$ and K$\protect\alpha _{2}$ radiations).}
\label{FigXray9K}
\end{figure}

To attempt to monitor the relative Na contents of the 22~K and 9~K phases we
studied the (008) x-ray Bragg reflections for the SC1 and SC2 samples. We found (see Fig.~\ref{FigXray9K}) that the SC2 sample displays a small splitting of the (008) reflection which, from our $c(x)$ calibration curve, would correspond to at most a 1\% increase in Na content from the 22~K to the 9~K phase.

\bibliography{NaxCoO2}

\end{document}